\documentclass{svjour2}                    
\smartqed  
\usepackage{graphicx}
\begin{document}

\title{Rolling down from the 30 keV peak
}
\subtitle{\\ Modelling the hard X--ray and $\gamma$--ray backgrounds}

\titlerunning{the hard XRB}        

\author{Andrea Comastri \and Roberto Gilli \and  G\"unther Hasinger}

\authorrunning{Comastri et al.} 

\institute{A. Comastri \at
              INAF--Osservatorio Astronomico di Bologna, Italy \\
              Tel.: +39-051-2095746\\
              Fax:  +39-051-2095700\\
              \email{andrea.comastri@bo.astro.it}           
           \and
            R. Gilli \at
              INAF--Osservatorio Astronomico di Bologna, Italy \\
              Tel.: +39-051-2095719\\
              Fax:  +39-051-2095700\\
              \email{roberto.gilli@bo.astro.it}           
           \and
	    G. Hasinger \at
	      Max Planck f\"uer Extraterrestrische Physik, Garching, 
		Germany  \\
	      Tel.: +49-89-300003401\\
	      Fax : +49-89-300003569 \\
              \email{ghasinger@mpe.mpg.de}           
}

\date{Received: date / Accepted: date}

\maketitle

\begin{abstract}

We will briefly discuss the importance of sensitive X--ray observations 
above a few tens of keV for a better understanding 
of the physical mechanisms associated to the Supermassive Black Hole 
primary emission in both radio quiet and radio loud AGN and
to the cosmological evolution of the most obscured sources.

\keywords{X--rays \and background radiation \and AGN}
\end{abstract}

\section{Introduction}
\label{intro}

The fraction of the hard X-ray background (XRB) resolved into
discrete sources by deep {\it Chandra} and XMM--{\it Newton} observations 
smoothly decreases from essentially 100 \% below 2--3 keV 
to about 50\% in the 7--10 keV energy range (Worsley et al. 2005).  
The resolved fraction averaged on the standard 2--10 keV band is
of the order of 80\% (see Hickox \& Markevitch 2006 for a recent estimate). 
At energies greater than 10 keV, where the bulk of the 
background energy density is produced, the resolved fraction is negligible, 
being strongly limited by the lack of imaging X--ray observations.
\par
The energy dependence of the resolved fraction goes almost 
hand in hand with the self--consistency  of AGN synthesis models. 
While at relatively low energy (say below 8--10 keV) a robust model,  
build over the AGN unified scheme, precisely account for a large
body of observational data (XRB spectral intensity, 
X--ray source counts, redshift distribution, etc.) at higher energies 
the predictive power is strongly limited 
by the lack of observational constraints. 
\par
At present the best estimates of the key parameters, 
responsible of the XRB spectral intensity around and above the 30 keV 
peak, mainly rely on the observations obtained with the {\em PDS} 
instrument onboard {\em BeppoSAX} (Risaliti et al. 1999; Matt 2004) and 
are thus limited to nearby bright ($S_{10-100 keV} > 10^{-11}$ erg cm$^{-2}$
s$^{-1}$) objects. 
As a consequence a relatively wide portion of the model parameter 
space remains so far unexplored (e.g. Comastri 2004a).

\section{XRB modelling}
\label{sec:1}


Within the framework of AGN synthesis models the key parameters 
responsible for the shape and intensity of the $>$ 10 keV 
XRB spectrum are:

\begin{itemize} 

\item the covering fraction and geometrical distribution of the 
cold dense gas responsible of the reflection "hump" peaking at 20--30 keV 
and observed in both type 1 and type 2 AGN.  

\item 
the relative fraction of heavily obscured sources with column densities
of the order of a few $10^{24}$ cm$^{-2}$ (the so called "mildly" 
Compton thick AGN; see Comastri 2004b for a review).  

\item 
the high energy cut--off of the primary emission which is usually 
parameterized as an exponential roll-over with an e--folding energy 
of a few hundreds of keV.

\end{itemize} 
 
\begin{figure*}
 \includegraphics[width=0.99\textwidth]{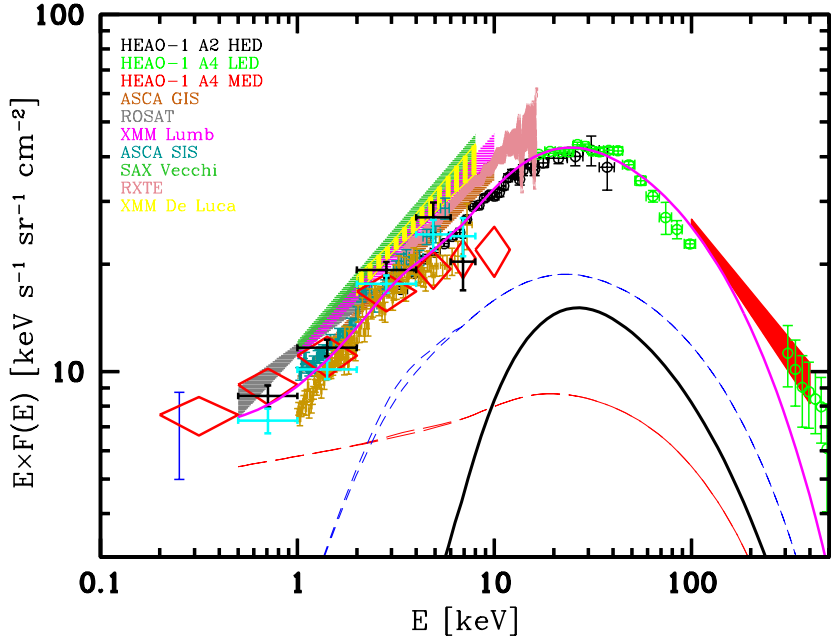}
\caption{The cosmic XRB spectrum and predicted AGN contribution 
(magenta solide line which includes also galaxy clusters)
splitted between unobscured (red dashed line), Compton thin 
(blue thin line) and Compton thick (black line). For unobscured and
Compton thin AGN the effect of including iron line emission is also 
reported and corresponds to the upper red and blue 
dashed curve (Gilli et al. 1999). 
The different XRB measurements are explained on the top left.  
Also shown are the resolved XRB fractions in different surveys 
by Worsley et al. (2005): Lockman Hole = red diamonds; 
CDFS = cyan crosses; CDFN = black crosses.} 
\label{fig:1}       
\end{figure*}

The reflection of hard X--rays by cold dense circumnuclear gas is a 
well established property of nearby Seyfert galaxies (see Pounds  
et al. 1990 for the discovery {\em Ginga} observations and Perola 
et al. 2002 for a spectral survey of bright AGN with {\em BeppoSAX}).
For unobscured or relatively unobscured AGN the reflecting material 
is covering about $2\pi$ sr. 

Several independent observational evidences suggest that 
a fraction as large as 50\% (or even higher) of the Seyfert 2 galaxies in the 
local Universe are obscured by Compton--thick material 
(Risaliti et al. 1999; Guainazzi et al. 2005; Comastri 2004b). 
A population of Compton thick (hereinafter CT) sources has to be included 
in AGN synthesis models for the XRB in order to match
the 30 keV intensity peak and at the same time to avoid 
an excessive number of Compton thin AGN which are not detected 
in {\it Chandra} and XMM--{\it Newton} deep surveys 
(see the discussion in Comastri 2004b). 
Their relative fraction and cosmological evolution are free parameters
which are tuned until a good fit to the XRB spectrum is obtained. 

Assuming the same cosmological evolution of soft X--ray selected 
type 1 AGN of Hasinger et al. (2005), the best fit model shown 
in Figure 1 is obtained with a relative fraction of CT AGN which 
is the same of Compton thin ($N_H$ in the range 10$^{22}$--10$^{24}$ 
cm$^{-2}$) and twice than unobscured ($N_H < 10^{22}$ cm$^{-2}$) AGN 
(see Gilli et al 2006 for further details).
The peaked shape of the CT spectrum in a $\nu F_{\nu} - \nu$ plot
is due to heavy absorption at low energies and the high energy cut--off 
($E_{cut}$). The latter, fixed at 320 keV in Fig.~1,  
has to be present in the high energy spectrum of all the sources 
(regardless of the absorption column density)  
otherwise the observed XRB flux above about 100 keV would be exceeded.

The high energy cut--off e--folding energy has been unambiguosly 
measured only for a bunch of nearby Seyfert galaxies. The best fit values 
are loosely constrained in the 150--350 keV range 
(Matt 2004; Malizia et al. 2003). Both BeppoSAX (Matt et al. 1999) and more 
recent {\em INTEGRAL} observations 
(Soldi et al. 2005) suggest a large scatter in the $E_{cut}$ values 
which span from about 50 keV up to lower limits of the order of 500 keV.  
While a clear bias against large values of the high energy cut--off 
is present, the available observations seem to indicate that
such a parameter might not be the same in all the sources at variance with 
the assumptions of most of the synthesis models.
\par



The properties of the sources of the "unresolved" background 
may be inferred by subtracting from the observed broad band 
spectrum the energy dependent fraction of the "resolved" XRB.
Such an approach has been pursued both making use of present 
observations (Worsley et al. 2005) and exploiting the 
XRB synthesis models (Comastri 2004).
In the former case the spectral shape  of the 
"unresolved" background  is consistent with that expected from a 
population of obscured 
($N_H \simeq 10^{23-24}$ cm$^{-2}$) AGN at redshifts $\sim$ 0.5--1.5.
In the latter, following a model dependent approach, 
the shape of the unresolved fraction can be predicted 
over a much wider energy range. 
More specifically assuming a set of model parameters 
tuned to match the 2--10 keV resolved fraction and to reproduce 
the observed absorption distribution in deep fields the predicted 
XRB flux and spectral shape above 10 keV  are strongly dependent 
from the adopted of $E_{cut}$ value (see Fig.~2 in Comastri 2004a).

\begin{figure*}
 \includegraphics[width=0.99\textwidth]{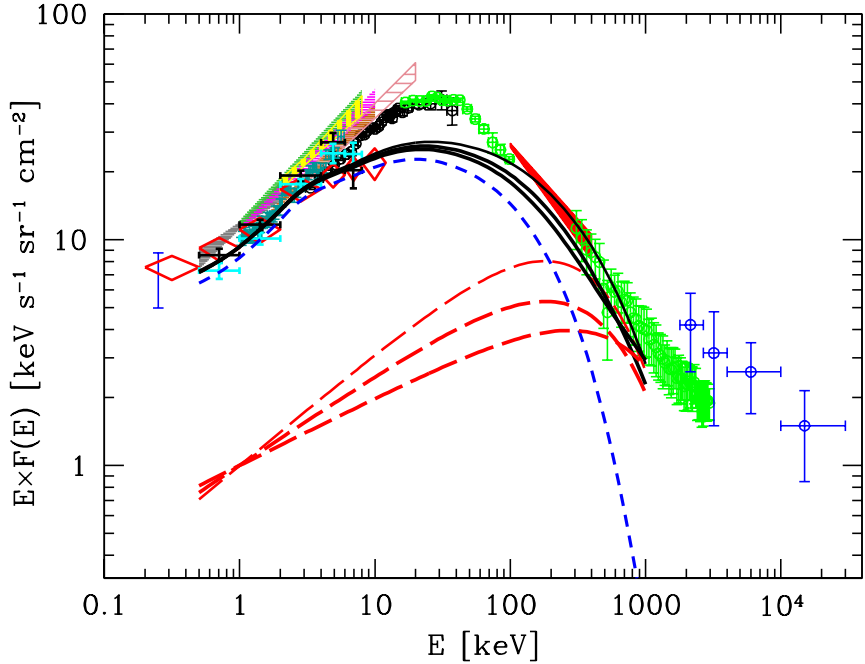}
\caption{An estimate of the radio loud AGN contribution to the XRB computed 
for a range of slopes and high energy cut-offs (red long-dashed lines; 
see details in the text). The blue short dashed line represents the so far 
resolved fraction modeled following the prescriptions of Gilli et al. 2006. 
The solid black lines are the sum of the above mentioned components}
\label{fig:2}       
\end{figure*}

\section{Exploring the parameter space}

An additional source of uncertainties comes from the possible
contribution to the hard XRB / soft $\gamma$--ray backgrounds 
(above about 100 keV) of the population of radio--loud sources 
which includes, for the purposes of the present exercise, 
both flat spectrum radio quasars (e.g. 3C 273) and BL Lac objects. 
We refer to Giommi et al. (2006 and these proceedings) for 
a detailed discussion of their contribution 
to the extragalactic X--ray and $\gamma$--ray bands.
In the very first approximation  radio--loud sources are expected
to provide the bulk of the $\gamma$--ray background above a few tens of MeV,
while radio--quiet AGN dominate below 100 keV. 
In between their relative contribution is expected to be comparable.
An attempt to estimate the  contribution of radio--loud 
quasars to the X--ray/soft $\gamma$--ray background 
(from 1 keV to a few MeV) is presented here. The assumed template spectrum 
and cosmological evolution are as follows:

\begin{itemize}

\item 
the average X--ray/$\gamma$--ray spectrum is parameterized with 
a single power law plus a high energy cut--off. 

\item 
the input spectrum is then folded with the same luminosity function
of radio--quiet AGN assuming a pure luminosity evolution of the form
$L(z) \propto L(0) \times (1+z)^3$ normalized to account for 10\% of the 
XRB flux at 1 keV in agreement with Giommi et al. (2006) and 
Galbiati et al. (2005).

\end{itemize}

The power law slope energy index $\alpha$ is varied in the range 
0.5--0.7 consistent with the averaged value observed in the X--ray band 
over a large range of redshifts (i.e.Reeves \& Turner 2000; 
Page et al. 2005; Lopez et al. 2006), while the choiche of the 
$E_{cut}$ value is driven by the requirement of not exceeding the 
background flux. The results for three representative pairs of 
$\alpha$ and $E_{cut}$ : (0.5 , 800 keV); (0.6 , 1 MeV) and 
(0.7 , 2 MeV) are shown in Figure 2 
(long dashed lines from top to bottom respectively). 
In our approach, almost by definition, the two parameters 
are not completely independent, the flatter the power law slope, the lower the
cut--off energy is. 
The contribution of radio loud AGN is then summed to the observationally 
resolved fraction in the 2--10 keV range,
extrapolated to high energies following the prescriptions of the
Gilli et al. (2006) model (short dashed line in Fig.~2). Finally
the total contribution for the three $\alpha$ , $E_{cut}$ pairs is 
computed (solid lines).
 
\begin{figure}
 \includegraphics[width=0.9\textwidth]{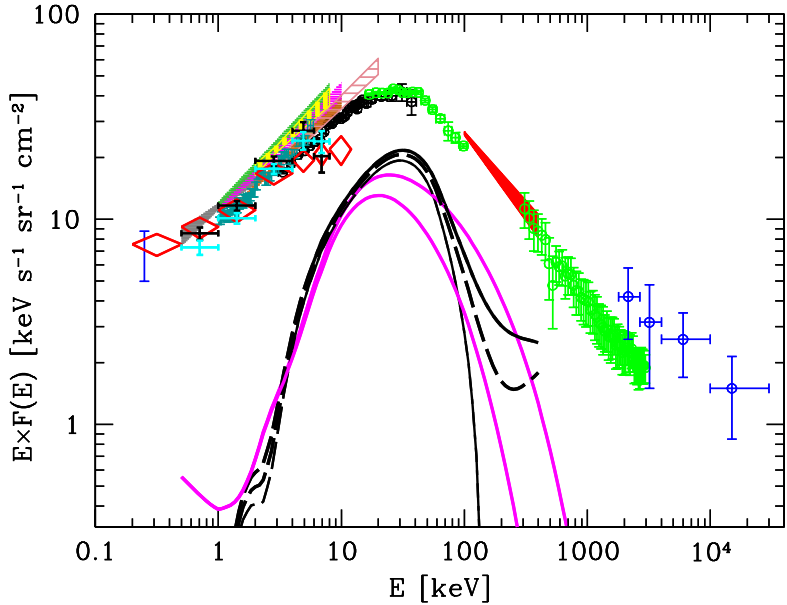}
\caption{The unresolved background as obtained by subtracting the contribution
of both radio--quiet and radio--loud AGN (long-dashed black lines) along 
with that predicted by the CT AGN in the Gilli et al. 2006 
model for two different $E_{cut}$ values (magenta solid lines). 
The XRB spectral data in the 0.3--3 MeV range are from {\rm SMM} 
(Watanabe et al. 1997), while in the 2--30 MeV from 
{\rm COMPTEL} (Kappadath et al. 1996)}
\label{fig:3}       
\end{figure}


The residual unresolved background obtained by subtracting the models
previously described from the total XRB flux, modeled using the 
Gruber (1992) analytical approximation, is shown in Figure 3.
A comparison with the contribution of a population of CT
AGN, assumed to evolve as unobscured sources (Hasinger et al. 2005),   
for two different values (100 and 320 keV) of the high energy cut--off 
is shown in Fig.~4. The model spectra agree extremely well below 
the peak energy while at higher energies, depending from the $E_{cut}$ and
$\alpha$ values adopted to model the contribution of radio--loud AGN, 
the predicted residual spectrum falls more rapidly with an 
over-exponential shape which is more pronounced for $\alpha$=0.5, 
$E_{cut}$=800 keV. 
At the face value these results suggest that the sources of the 
unresolved background are characterized by either a very peaked 
spectrum or are clustered over a smaller redshift range or a 
combination of the two.

\section{Conclusions} 

The above described exercise confirm and somewhat extend the findings
of Worsley et al (2005) and Comastri (2004a).
Even though the model dependent approach does not allow
to break the degeneracy between the spectral parameters and in particular
among the cut--off energy and the source redshift some preliminary 
conclusions may be already drawn.
A population of CT AGN with a space density and cosmological evolution 
comparable to that of less obscured,Compton thin sources 
provide an excellent description of the residual XRB spectrum up to 
about 10--20 keV. 
At higher energies the residual spectrum is no longer well approximated
by an exponentially cut-offed power law model folded with the redshift 
distribution observed in deep {\it Chandra} and XMM--{\it Newton} fields.
It can be argued that a suitable fine tuning of cut--off energy and redshift 
distributions along with a more extended treatement of the contribution 
of radio--loud AGN may provide a much better fit to the unresolved 
XRB above 30 keV. 
However it should be also stressed that the observational constraints
used to compute the residual XRB are still subject to significant
uncertainties. Recent measurements of the XRB spectral intensity 
below 10 keV obtained with imaging instrument are systematically 
higher than the intensity originally measured by HEAO1--A2 (see Fig.~1 and
Revnivtsev et al. 2005 for a recent reanalysis of the HEAO1 data), 
Given that the the only available observation of the 
intensity and location of the XRB peak is from the same A2 experiment, 
some doubts have been raised on the absolute flux and intensity of the
XRB above 10 keV.
In the hard X--rays/ soft $\gamma$--rays the statistical errors as well
as the relative calibration between different experiments 
are even larger. 
Before attempting a more sophisticated modelling of the broad band background
spectrum from its peak down to MeV and GeV region a more robust determination 
of the observational framework is needed. 
Sensitive imaging  observations down to about 10$^{-14}$ erg cm$^{-2}$ 
s$^{-1}$ in the approximately 10--50 keV energy range would resolve about 
half of the background in that band to be compared with the present 
less than a few percent. A major breaktrough in the census and the study 
of obscured accreting black holes is expected by future 
hard X--ray missions and especially by Simbol-X (Ferrando et al. 2005).
A mission capable to explore the 100--1000 keV decade with 
comparably good sensitivity (i.e.the Gamma Ray Imager
mission concept see Von Ballmoos and Kn\"odlseder these proceedings) 
would open the possibility to investigate the origin of the 
primary emission mechanism in AGN, presumably responsible of the high 
energy cut--off, with important consequences for the scientific objectives
discussed in this paper.

\section{Acknowledgements} 

This work was partially funded by a MIUR grant (Cofin 03-02-23) and by ASI.
AC thanks Gabriele Ghisellini for useful discussions, Filippo Frontera 
for the kind invitation and Peter Von Ballmoos for organizing an interesting 
and stimulating workshop. 

%
%



\begin{thebibliography}{}
%
%
\bibitem{Ref1}
Comastri A., in Multiwavelength AGN Surveys; proceedings of the 
Guillermo Haro Conference held December 8-12, 2003, in Cozumel, Mexico. 
Edited by R. Mujica and R. Maiolino. World Scientific Publishing Company, 
Singapore, p.323, (2004a) 
\bibitem{Ref2}
Comastri A.  Compton-Thick AGN: The Dark Side of the X-Ray Background 
in Supermassive Black Holes in the Distant Universe. 
Edited by A. J. Barger, Kluwer Academic Publishers, 
Dordrecht, The Netherlands, 2004, p.245 (2004b)
\bibitem{RefPh}
Ferrando P., et al. SIMBOL-X: a formation flying mission for hard X--ray 
astrophysics   Optics for EUV, X-Ray, and Gamma-Ray Astronomy II. 
Edited by Citterio, O. and O'Dell S.L. Proceedings of the SPIE, 
Vol. 5900, p. 195 (2005).
\bibitem{refga}
Galbiati E., Caccianiga A., Maccacaro T., Braito V., 
Della Ceca R., Severgnini P., Brunner H., Lehmann I., Page M.J. :	
XMM--{\it Newton} spectroscopy of an X--ray selected sample of RL AGNs,
A\&A 430, 927 (2005)
\bibitem{refGi99}
Gilli R., Comastri A., Brunetti G., Setti G.: The contribution of AGN to 
the X-ray background: the effect of iron features New Astronomy 4, 45 (1999)
\bibitem{refGi}
Gilli R., Comastri A., Hasinger G. in preparation (2006). 
\bibitem{refG}
Giommi P., Colafrancesco S., Cavazzuti E., Perri M., Pittori C. : non--thermal 
cosmic backgrounds from blazars: the contribution to the CMB, X--ray and 
$\gamma$--ray backgrounds, A\&A 445, 843 (2006)
\bibitem{RefGr}
Gruber D.E. The Hard X-Ray Background in: The X-ray Background. 
Collected Papers and Reviews from a Workshop held in Laredo, Spain, 
September, 1990., X. Barcons, A.C. Fabian eds., Cambridge University Press, 
Cambridge; New York, NY, 1992. p. 44 (1992) 
\bibitem{RefGu}
Guainazzi M., Matt G., Perola G.C. X-ray obscuration and obscured AGN in 
the local universe A\&A 444, 119 (2005)
\bibitem{RefHa}
Hasinger G., Miyaji T, Schmidt M.  Luminosity-dependent evolution of soft 
X-ray selected AGN: New {\it Chandra} and XMM--{\it Newton} surveys 
 A\&A  441, 417 (2005) 
\bibitem{RefH}
Hickox R.C., Markevitch M., Absolute measurement of the unresolved 
X--ray background in the 0.5-8 keV band with Chandra ApJ in press 
(astro-ph/0512542) (2006)
\bibitem{RefK}
Kappadath S.C. et al. : The preliminary cosmic diffuse $\gamma$--ray 
spectrum from 800 keV to 30 MeV measured with COMPTEL A\&AS 120, 619 (1996)
\bibitem{RefL}
Lopez L.A., Brandt W.N., Vignali C., Schneider D.P., Chartas G., Garmire G.P.:
A Chandra snapshot survey of representative high redshift radio--loud 
quasars from the Parkes--MIT--NRAO sample, AJ in press astro--ph/0601037 (2006)
\bibitem{RefMa}
Matt G. et al. One more surprise from the Circinus Galaxy: 
BeppoSAX discovery of a transmission component in hard X-rays A\&A 341 L39 
(1999) 
\bibitem{RefC}
Matt G.: Broad band spectral properties of AGNs and quasars, observations and theory
Nuclear Physics B -- Proceedings Supplements Vol. 132, p. 97 (2004) 
\bibitem{RefD}
Malizia A., Bassani L., Stephen J.B., Di Cocco G., 
Fiore F., Dean, A.J. BeppoSAX Average Spectra of Seyfert Galaxies ApJ 589, L17 
(2003) 
\bibitem{RefPa}
Page K.L., Reeves J.N., O'Brien P.T., Turner M.J.L.: XMM--{\it Newton} 
spectroscopy of high-redshift quasars  MNRAS 364, 195 (2005) 
\bibitem{RefPe}
Perola G.C., Matt G., Cappi M., Fiore F., Guainazzi M., 
Maraschi L., Petrucci P.O., Piro, L. Compton reflection and iron 
fluorescence in BeppoSAX observations of Seyfert type 1 galaxies
A\&A 389, 802 (2002)
\bibitem{RefPo}
Pounds K.A., Nandra K., Stewart G.C., George I.M., Fabian A.C.	
X-ray reflection from cold matter in the nuclei of active galaxies
Nature 344, 132 (1990)
\bibitem{RefB}
Risaliti G., Maiolino R., Salvati M., The distribution of absorbing column 
densities among Seyfert 2 galaxies, ApJ 522, 157 (1999)
\bibitem{RefRe}
Reeves J.N., Turner M.J.L. X-ray spectra of a large sample of quasars with ASCA
MNRAS 316, 234 (2000)
\bibitem{RefE}
Revnivtsev M., Gilfanov M., Jahoda K., Sunyaev R.	
Intensity of the cosmic X--ray background from HEAO1/A2 experiment
A\&A 444, 381 (2005) 
\bibitem{RefE}
Soldi S., Beckmann V., Bassani L., Courvoisier T.J.L., Landi R., 
Malizia A., Dean A.J., de Rosa A., Fabian A.C., Walter R. :	
INTEGRAL observations of six AGN in the Galactic Plane A\&A 444, 431 (2005)
\bibitem{RefW}
Watanabe K., Hartmann D.H., Leising M.D., The L.-S., Share G.H., Kinzer R.L.
The Cosmic Gamma Ray Background from Supernovae in 
Proceedings of the Fourth Compton Symposium, Eds. C. D. Dermer, 
M. S. Strickman, J. D. Kurfess, Williamsburg, VA April 1997: 
AIP Conference Proceedings 410, p. 1223. (1997) 
\bibitem{RefWo}
Worsley M.A., Fabian A.C., Bauer F.E., Alexander D.M., Hasinger G., 
Mateos S., Brunner H., Brandt W.N., Schneider D.P. The unresolved hard 
X-ray background: the missing source population implied by the 
{\it Chandra} and XMM--{\it Newton} deep fields MNRAS 357, 1281 (2005)
\end{thebibliography}


\end{document}